\documentclass[11pt,a4paper]{article}   
\usepackage{amsmath}
\usepackage{amssymb}
\usepackage{mathrsfs}
\usepackage{wrapfig}

\usepackage{rotating}
\usepackage{color}  

\usepackage[round]{natbib}

\usepackage{fancybox}

\usepackage{xcolor}
\usepackage[normalem]{ulem}
\usepackage{graphicx}

\usepackage{marginnote}

\usepackage{xparse}

\NewDocumentCommand{\lumar}{mo}
 {
  \IfValueTF{#2}
  {\marginnote[\parbox{45pt}{#1}$\;\Rightarrow$]{$\Leftarrow\;$\parbox{75pt}{#1}}[#2] }
  {\marginnote[\parbox{45pt}{#1}$\;\Rightarrow$]{$\Leftarrow\;$\parbox{75pt}{#1}}[-20pt] }
 }

\newcommand{\ie}{i.e.\ }

\begin{document}
\title{An idealised experimental model of ocean surface wave transmission by an ice floe}
\author{L.~G. Bennetts$^{1}$, A. Alberello$^{2}$, M. H. Meylan$^{3}$, C. Cavaliere$^{4}$, A.~V. Babanin$^{2}$ and A. Toffoli$^{2}$
\\
{\footnotesize
$^{1}$School of Mathematical Sciences, University of Adelaide, Adelaide 5005, Australia}
\\
{\footnotesize
$^{2}$Centre for Ocean Engineering Science and Technology, Swinburne University of Technology, Hawthorn 3122, Australia}
\\
{\footnotesize
$^{3}$School of Mathematical and Physical Sciences, University of Newcastle, Callaghan 2308, Australia}
\\
{\footnotesize
$^{4}$Polytechnic University of Milan, Milan 20133, Italy}
}
\date{\today}
\maketitle

\begin{abstract}
An experimental model of transmission of ocean waves by an ice floe is presented. 
Thin plastic plates with different material properties and thicknesses are used to model the floe. 
Regular incident waves with different periods and steepnesses are used, ranging from gently-sloping to storm-like conditions. 
{A wave gauge is} used to measure {the} water surface elevation in the lee of the floe. 
The depth of wave overwash on the floe is measured by a gauge in the centre of the floe's upper surface. 
Results show transmitted waves are regular for gently-sloping incident waves but irregular for storm-like incident waves. 
The proportion of the incident wave transmitted is shown to decrease as incident wave steepness increases,
and to be at its minimum for an incident wavelength equal to the floe length.
Further,  
{a trend is noted for transmission to decrease as the mean wave height in the overwash region increases.}
\end{abstract}

\section{Introduction}

Ocean surface waves penetrate up to hundreds of kilometres into the sea ice-covered ocean \citep[e.g.][]{Squ&Mor80,Kohetal14}.
The region occupied by waves is known as the marginal ice zone (\textsc{miz}).
Waves have a profound impact on the ice cover in the \textsc{miz}.
They (i) fracture large floes into smaller, more mobile and vulnerable floes \citep{Lanetal98}, 
(ii) herd floes \citep{Wad83},
(iii) introduce warm water and overwash the floes, thus accelerating ice melt \citep{Wadetal79a,Mas&Sta10},
and (iv) cause the floes to collide, which erodes the floes and influences the large-scale deformation of the ice field via momentum transfer \citep{Sheetal87,Mar&Bec87,Mar&Bec88}.
\cite{Kohetal14} recently identified a negative correlation between trends in local wave activity and trends in regional ice extent in the Antarctic \textsc{miz},
which is conjectured to result from impacts of large-amplitude storm waves on the ice cover.

Interactions between waves and the ice cover cause wave energy to reduce with distance travelled into the \textsc{miz}.
Moreover, the ice cover reduces the energy of short-period waves more rapidly than long{er}-period waves \citep{Wadetal88a,Meyetal14}.
Incident wave spectra, therefore, skew towards long periods as they propagate deeper into the \textsc{miz}, 
in addition to a reduction of energy held by each spectral component.

The ice cover in the \textsc{miz} is composed of ice floes with diameters on the order of metres to hundreds of metres.
The ratio of the prevailing floe diameters to the incident wavelengths determines the form of the wave-ice interactions,
and, hence, the mechanisms responsible for reducing wave energy.
In particular, waves perceive a field of floes with diameters much smaller than their wavelength, for example, pancake ice,
as an effective medium.
In this regime, wave energy is reduced with penetration distance due to viscous losses \citep[e.g.][]{Kel98,DeC&Des02,Wan&She11,Zha&She13}.

In contrast, waves distinguish individual floes with diameters comparable to their wavelength,
for example, floes produced by wave-induced fracture.
The individual floes reflect a proportion of the incident wave {energy}, dissipate a proportion and transmit the remaining proportion.
Theoretical and numerical methods have been developed to predict the rate at which wave energy reduces in a large field of floes
by combining transmission properties of individual floes \citep[e.g.][]{Mas&LeB89,Meyetal97,Benetal10,Ben&Squ12b,Ben&Squ12a}.
The models have recently been used to parameterise wave propagation in the \textsc{miz} in large-scale ocean wave and sea ice models
\citep{Dob&Bid13,Wiletal13a,Wiletal13b}. 

The present investigation focusses on modelling transmission of waves by a {solitary} ice floe.
A sum of processes involved in wave-floe interactions determine the proportion of wave energy transmitted.
Floes bend and flex in response to wave motion, in addition to responding in the standard six rigid-body degrees of motion.
{Waves experience drag (form and skin) travelling around the floe \citep{Kohetal11}.}
The floes are also susceptible to drift.
Further, 
{t}he small freeboards of floes permit waves of moderate amplitude to overwash the floes, 
and the shallow draughts permit energetic waves to slam floes against the ocean surface.

A series of mathematical models of wave interactions with an ice floe, and hence transmission, 
have been developed \cite[e.g.][]{Mas&LeB89,Mey02,Ben&Wil10}.
Thin plates are used to model the floes. 
{T}he models are conservative, \ie no energy dissipation{,
and assume all motions are proportional to the incident wave amplitude, \ie the proportion of wave energy transmission does not depend on the incident amplitude.} 
They neglect wave overwash of the floes by waves, slamming{, drag} and drift.
 
An idealised experimental model of wave transmission by an ice floe is presented here.
Consistent with the mathematical models, thin plastic plates were used to model the floe.
Regular (monochromatic) incident waves, ranging from gently-sloping to storm-like, were used.
The transmitted wave field was measured by {a wave gauge}.

\citet{Kohetal07}, \citet{Huaetal11}, \citet{Monetal13a,Monetal13b} and \citet{McG&Bai14} 
recently used closely related experimental models to investigate wave-induced motions of ice floes.
However, the present investigation is the first to study transmission of waves by an ice floe.

{Field measurements of wave energy at discrete points in the \textsc{miz} exist \citep[e.g.][]{Squ&Mor80,Wadetal88a,Meyetal14}.
Wave transmission by a large collection of floes can be inferred from these measurements.
However, the measurements are not accompanied by detailed information of floe properties.
Field measurements of transmission of waves by an individual floe of known properties do not exist.}

{The experimental model is used to gain understanding of how a floe affects wave propagation,
with respect to floe and incident wave properties.
The study focusses on the proportion of wave energy transmitted by the floe, \ie the transmission coefficient.
The effect of the floes on the full wave spectrum is also analysed for a subset of the tests conducted.
Further, the transmission is related to the properties of the overwash, which were measured by a specifically designed wave gauge mounted on the model floe.}


\section{Experimental model}

\begin{wrapfigure}{r}{0.55\textwidth}
  \centering
  \noindent\includegraphics[width=0.5\textwidth]{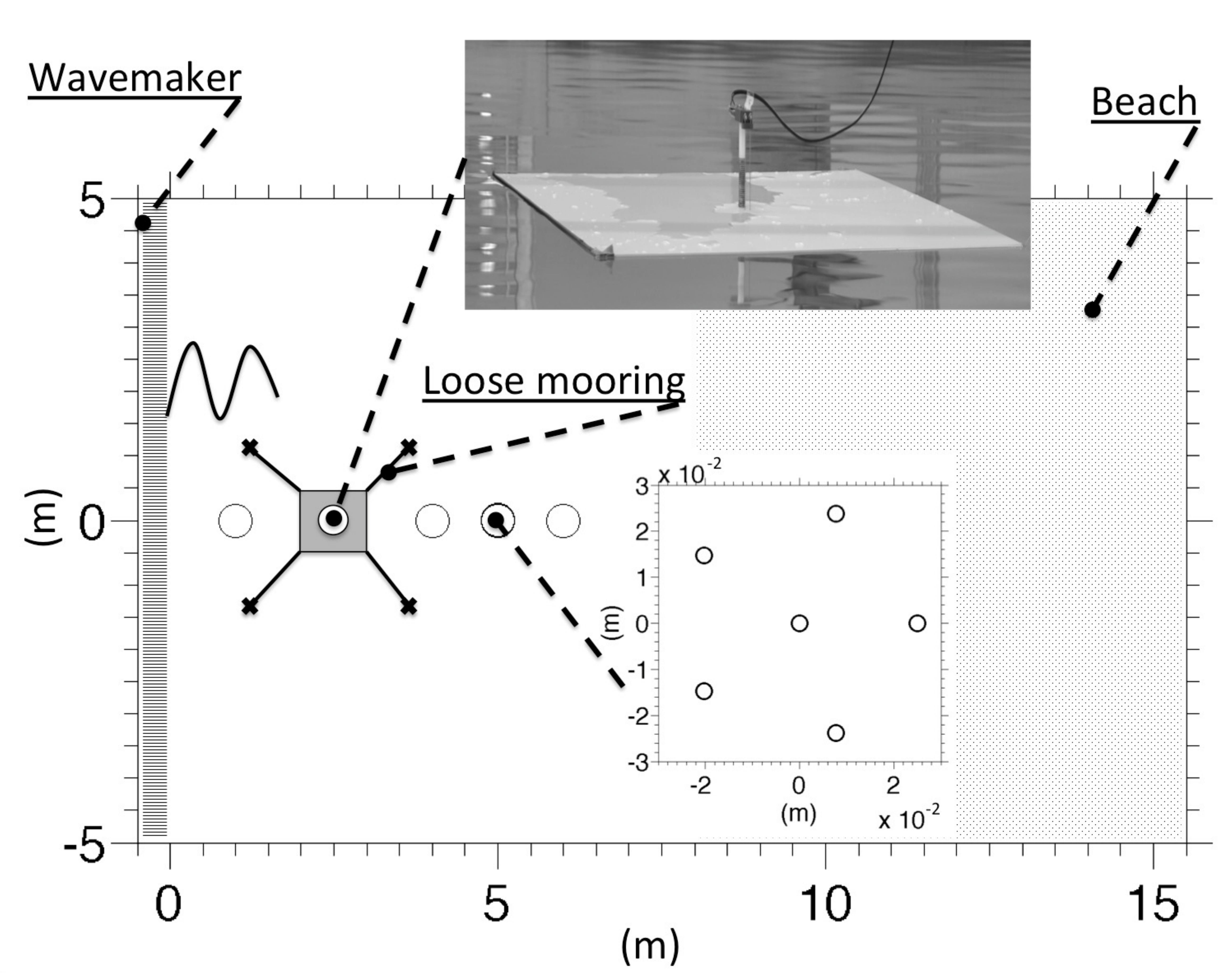}
  \caption{Schematic of the wave basin and experimental set-up. Circles denote locations of the gauges. Top inset is a photo of the model floe, which shows the gauge used to measure overwash. 
  Bottom inset shows the hexagonal arrangement of gauges in the lee of the floe.}\label{fig01}
\end{wrapfigure}

The experimental model was implemented in the wave basin at the Coastal Ocean And Sediment Transport (\textsc{coast}) 
laboratories of Plymouth University, U.K.
Fig.~\ref{fig01} shows a schematic plan view of the wave basin and experimental set-up.
The basin is 10\,m wide, 15.5\,m long and was filled with {fresh} water $H=0.5$\,m in depth. 
{The room and water temperatures were approximately 20$^{\circ}$\,C and 16$^{\circ}$\,C, respectively.}

At the left-hand end of the basin, a wave maker, consisting of twenty individually controlled active pistons, generate{d} incident waves. 
The pistons automatically adjust{ed} their velocities to absorb waves reflected by the basin walls or the floe. 
At the right-hand end of the basin, wave energy {was} dissipated by a beach with a linear profile of slope 1:10. 

The use of active pistons and a beach {could not} eliminate the presence of reflected waves in the basin perfectly.
However, a reflection analysis in the centre of the basin, from control tests conducted without a floe, 
showed the reflected energy to be less than 1\% of the incident wave energy. 
(Reflection is largest for low-frequency waves.)
Waves reflected by the beach and wave maker are, therefore, considered to have a negligible impact on the transmission measured in the tests.

A thin plastic plate was deployed 2\,m from the wave maker, as a model ice floe. 
Two different types of plastic were tested: 
polyvinyl chloride (\textsc{pvc}) with density $500$\,kg\,m$^{-3}$ 
and Young's modulus $500$\,MPa; and
polypropylene with density of $905$\,kg\,m$^{-3}$  
and Young's modulus $1600$\,MPa.

{A} sea ice density of  $900$\,kg\,m$^{-3}$ 
and Young's modulus of 6\,GPa are commonly assumed for wave-ice interactions.
{\citet{Tim&Wee10} review the mechanical properties of sea ice. 
They report density measurements from 720\,kg\,m$^{-3}$ to 940\,kg\,m$^{-3}$ with an average of 910\,kg\,m$^{-3}$, and Young's modulus measurements from approximately 1\,GPa to 10\,GPa.}
 
{For a specified laboratory-to-field geometric scaling factor,
the Young's modulus of the laboratory model floe is scaled by this factor to give its equivalent field value,
whereas the density is equivalent to its field value without scaling \citep{Tim80}. 
The polypropylene floes, therefore, have a more realistic density than the \textsc{pvc} floes,
whereas the \textsc{pvc} floes typically have a more realistic Young's modulus than the polypropylene floes.}

\textsc{Pvc} plates were provided with thicknesses $h=5$\,mm, 10\,mm and 19\,mm.
Polypropylene plates were provided with thicknesses $h=5$\,mm, 10\,mm and $20$\,mm. 
The plates were cut into square floes with side lengths $L_{floe}=1$\,m. 

Regular incident waves were generated, using three wave periods $T=0.6$\,s, 0.8\,s, and 1\,s, 
and corresponding wavelengths $L_{wave}=$0.56\,m, 1{.00}\,m and 
{1.51\,m}, 
{to two decimal places,} 
respectively. 
The tests, therefore, covered conditions in which incident waves were shorter than, equal to and longer than the floe. 
The wave amplitude, $a$, was selected so that the wave steepness {(a nondimensional form of wave amplitude)}, $ka$, where $k$ is the wavenumber, 
matched the following values: 0.04, 0.08, 0.1 and 0.15. 
This range includes gently-sloping waves ($ka=0.04$ and $0.08$) and storm-like waves ($ka = 0.1$ and 0.15), without reaching the breaking limit \citep{Tofetal10}. 
{The relative water depth, $kH$, is greater than 2 for all incident waves.
Wave interactions with the basin floor are, therefore, marginal, and the waves are considered to be in a deep water regime \citep{Tofetal09}.}

The water surface elevation, $\eta$, was monitored with capacitance gauges at a sampling frequency of 128\,Hz
{and an accuracy of approximately 0.1\,mm.}  
One gauge was deployed approximately 1\,m in front of the floe to capture the incident (and reflected) waves. 
In the lee of the floe, three gauges were deployed every metre to track the evolution of the transmitted wave field. 
The results showed the transmitted wave heights increase slightly with distance away from the floe, which is attributed to diffraction.
Results presented in \S~\ref{sec:res} are, therefore, taken from the gauge closest to the floe only.
At 2\,m from the rear edge of the floe a six-gauge array, arranged as a pentagon of radius of 0.25\,m, 
was deployed to monitor the directional properties of the wave field (not reported here). 
For each configuration, five-minute time series were recorded.

A small gauge was also deployed in the middle of the upper surface of the floe to measure the depth of overwashed fluid (see photo inset of Fig.~\ref{fig01}). 
A motion sensor with sampling frequency of 200\,Hz was deployed on top of the this gauge to monitor the motion of the floe. 
The combined mass of the gauge and motion sensor was negligible in comparison to the mass of the floe.
A loose mooring was applied at the four corners of the plate to avoid collisions with the wave gauges, \ie free drift was not permitted. 


\section{Results and analysis}\label{sec:res}


\subsection{Detailed results for 5\,mm thickness floes} \label{sec:res_5mm}

\begin{figure}
  \centering
  \noindent\includegraphics[width=0.8\textwidth]{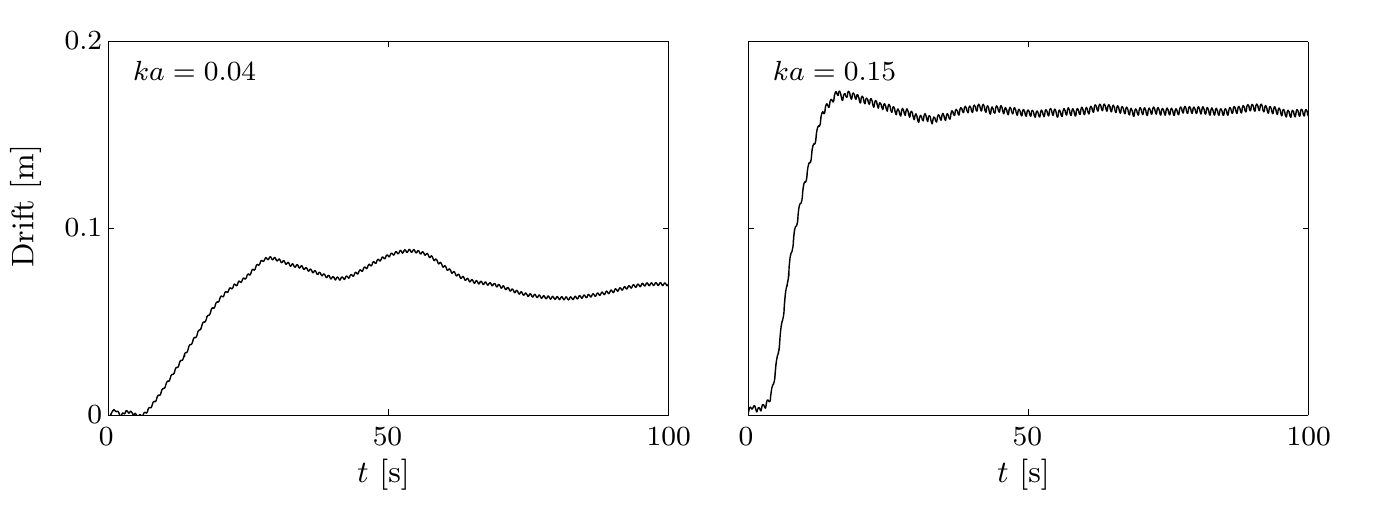}
  \caption{Distance drifted by a polypropylene floe of 5\,mm thickness as a function of time.
  Nondimensional incident wavelength $L_{wave}/L_{floe}=1{.00}$ and steepness $ka=0.04$ (left-hand panel) and $0.15$ (right).}
\label{drift}
\end{figure}

\begin{figure}
 \centering
  \noindent\includegraphics[width=0.9\textwidth]{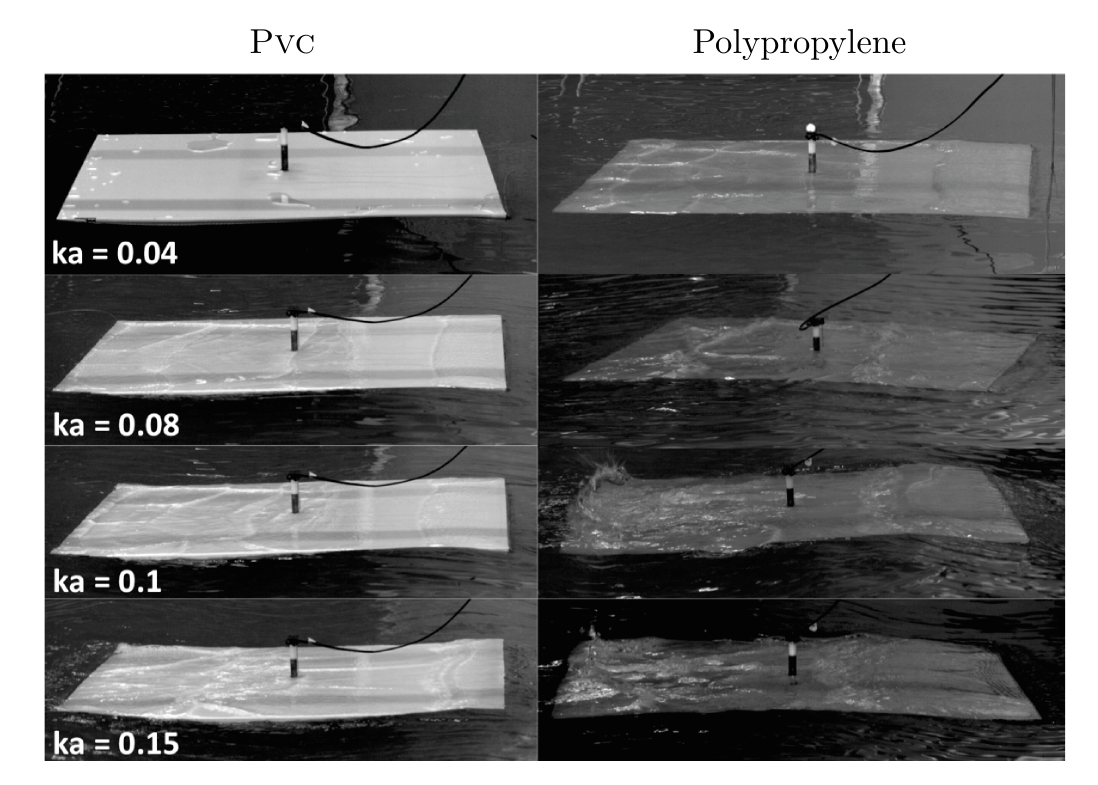}  
  \caption{Snapshots of 5\,mm thick floes for nondimensional incident wavelength $L_{wave}/L_{floe}=1{.00}$.}\label{photos}
\end{figure}

\begin{figure}
  \centering
  \noindent\includegraphics[width=0.7\textwidth]{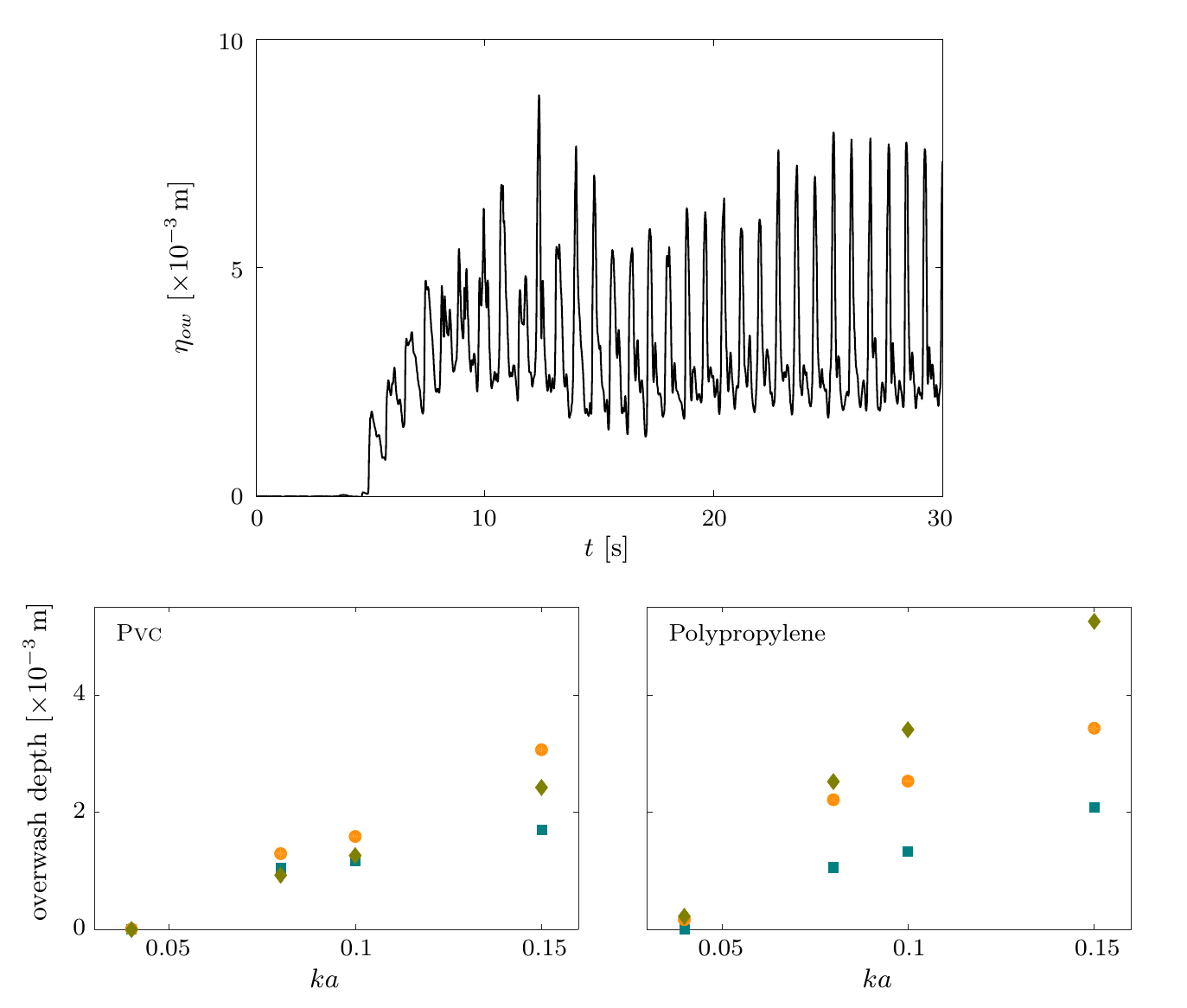} 
  \caption{Surface elevation at the centre of 5\,mm thick polypropylene floe  for nondimensional incident wavelength $L_{wave}/L_{floe}=1{.00}$\,m, and steepness $ka=0.15$ (top panel).
  {Mean o}verwash depth as a function of incident wave steepness (bottom panels),
  for nondimensional incident wavelengths $L_{wave}/L_{floe}=0.56$ (teal, squares), 1.00 (orange, circles) and 
  {1.51} (olive, diamonds).}
  \label{overwash}
\end{figure}

Detailed results are given here for the thinnest floes, with thickness 5\,mm.
The behaviours shown are broadly representative of the thicker floes tested. 

Fig.~\ref{drift} shows the wave-induced drift of the polypropylene floe for an incident wavelength, {nondimensionlised} with respect to the floe length, of $L_{wave}/L_{floe}=1{.00}$, 
and the most gentle steepness, $ka=0.04$ (left-hand panel), 
and most storm-like steepness, $ka=0.15$ (right). 
In both cases, the floe undergoes an oscillatory surge motion of period approximately equal to the incident wave period. 
The floe also experiences Stokes drift of approximately 0.001\,m\,s$^{-1}$ for the gently-sloping incident waves and 0.01\,m\,s$^{-1}$ for the storm-like incident waves \citep{Cha84}. 

The drift continues until the mooring is engaged. 
The floe then oscillates around a mean location.
The steeper incident wave applies a stronger drift force, which results in a greater tension in the mooring system and, hence, the mean location of the floe is almost twice as far displaced from the original floe location than for the more gentle incident wave.
Moreover, the mean location is attained far quicker for the steep incident wave.

Fig.~\ref{photos} shows example snapshots of the \textsc{pvc} and polypropylene floes under wave forcing.
The floes are submerged by the incident waves, either partially for the \textsc{pvc} floe, which has the larger freeboard, 
and/or gently-sloping incident waves,
or fully for the polypropylene floe and storm-like incident waves. 
{Note that floe rotation (yaw) is visible in some photos, in particular for the polypropylene floe and $ka=0.08$. 
Yaw is caused by small asymmetries in the incident waves, the initial floe alignment and the mooring system. 
It had a much longer period than the incident wave period.}

Overwash is generated at the front and rear ends of the floes, alternately as they pitch. 
The generation of overwash produces high-frequency free-wave components in the open water around the floe.
{S}hallow-water waves propagate in the overwashed fluid itself. 
When shallow-water waves, travelling in opposite directions, up and down the floes, meet, waves often become large enough to break, which dissipates wave energy.

The top panel of Fig.~\ref{overwash} shows an example time series provided by the  wave gauge on the surface of the polypropylene floe, $\eta_{ow}$, 
for the steepest incident wave, $ka=0.15$, and nondimensional wavelength $L_{wave}/L_{floe}=1{.00}$.  
{The overwash depth oscillates between approximately 2\,mm and 8\,mm here. 
The incident amplitude is approximately 24\,mm, \ie three times larger than the maximum overwash depth.
In general, the incident amplitude was an order of magnitude greater than the mean overwash depth.}

The bottom panels show {mean} overwash depth, as a function of incident wave steepness,
and organised according to nondimensional incident wavelength.
The results confirm overwash is deeper for the polypropylene floe and steeper incident waves.
For the polypropylene floe, overwash depth increases as the incident wavelength increases.
However, for the \textsc{pvc} floe, overwash is deepest for an incident wavelength equal to the floe length.

The most energetic storm-like waves slam the edges of the floe against the water surface. 
This contributes to generating additional high-frequency free-wave components in the surrounding open water.
Slamming was not measured.
However, it was visibly stronger for the denser polypropylene floe.

\begin{figure}
  \centering
  \noindent\includegraphics[width=0.8\textwidth]{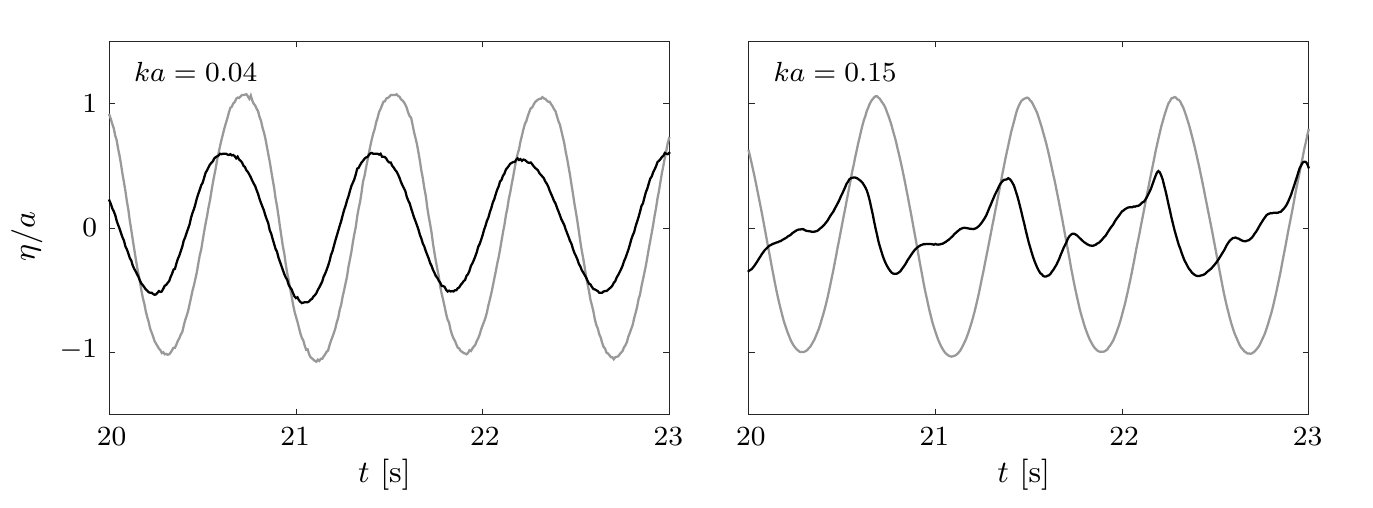}
  \caption{Surface elevation, $\eta$, in the lee of a 5\,mm thick polypropylene floe (thick black curves) and incident wave (grey), 
  normalised with respect to {target incident amplitude, $a$.} 
  Nondimensional incident wavelength $L_{wave}/L_{floe}=1{.00}$, and steepness $ka=0.04$ (left-hand panel) and $0.15$ (right).
  }\label{elevation}
\end{figure}

Fig.~\ref{elevation} shows example surface-elevation time series from the lee of the polypropylene floe. 
The surface elevations are 
{scaled by the target incident amplitudes, $a$.}
Incident waves were gently-sloping ($ka=0.04$, left-hand panel) 
and storm-like ($ka=0.15$, right).
Corresponding time series from control tests, conducted without the floe, are also shown. 

In both cases, the floe transmits only a proportion of the incident wave.
For the gently-sloping incident waves, the transmitted waves retain a regular profile.
For the storm-like incident waves, the transmitted waves are irregular,  
which is attributed to the strong overwash and slamming described above.

\begin{wrapfigure}{r}{0.6\textwidth}
  \centering
  \noindent\includegraphics[width=0.6\textwidth]{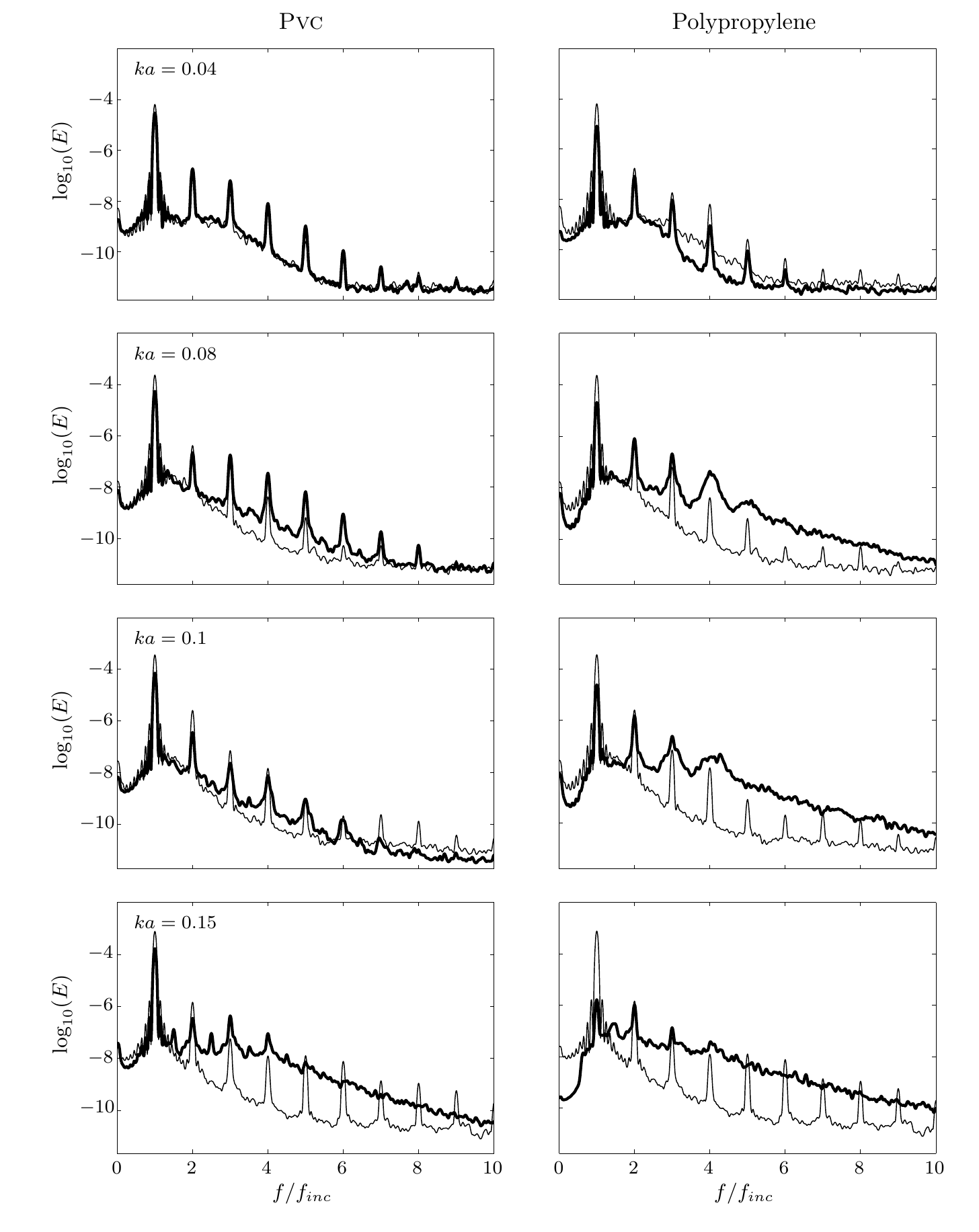}
  \caption{Wave spectra without floe (thin lines) and with 5\,mm thick floes (thick lines), for a nondimensional incident wavelength $L_{wave}/L_{floe}=1{.00}$.}
  \label{spec}
\end{wrapfigure}
 
Fig.~\ref{spec} shows the wave energy spectra, $E(f)$, where $f$ denotes frequency, in the lee of the floes. 
Corresponding wave spectra without the floe, from control tests, are also shown. 
The spectra are computed from segments of 4096 consecutive records and averaged over the entire time series. 

The wave spectra without the floe are dominated by the fundamental harmonic, 
\ie the frequency of the monochromatic wave generated.  
Higher-order harmonics, i.e. phase-locked bound waves, are also evident. 

For the smallest wave steepness tested, $ka=0.04$, the
polypropylene floe reduces the wave energy held in the fundamental harmonic {by approximately 90\%. It also reduces}  
the high-frequency energy. 
As the incident wave becomes steeper, the proportion of transmitted wave energy in the fundamental harmonic becomes increasingly damped by the polypropylene floe.  
Further, for the three largest steepnesses, the high-frequency tails of the spectra transmitted by the polypropylene floe contain 
greater energy than the incident spectra, and the higher-order harmonics are smeared out. 

The \textsc{pvc} floe also increasingly damps the fundamental harmonic, 
and transfers energy to the high-frequency tail, as the incident wave becomes steeper.  
However, the effects are weaker than for the polypropylene floe{. For example, the \textsc{pvc} floe reduces the energy 
held in the fundamental harmonic by approximately 50\% for $ka=0.04$ and 80\% for $ka=0.15$, which are obscured by the logarithmic scales used for Fig.~\ref{spec}.}
{The weaker impact of the \textsc{pvc} floe than the polypropylene floe supports} 
the hypothesis {that} overwash and slamming {affect transmission}.

The proportions of wave energy transmitted by the floes, \ie transmission coefficients, 
are quantified in terms of mean wave heights,
due to the irregular behaviour of the transmitted waves for storm-like incident waves.
{Mean} wave heights are calculated via the zero down-crossing method, 
and subsequent averaging over the time series.

\begin{figure}
  \centering
  \begin{tabular}{c c}
   \noindent\includegraphics[width=0.8\textwidth]{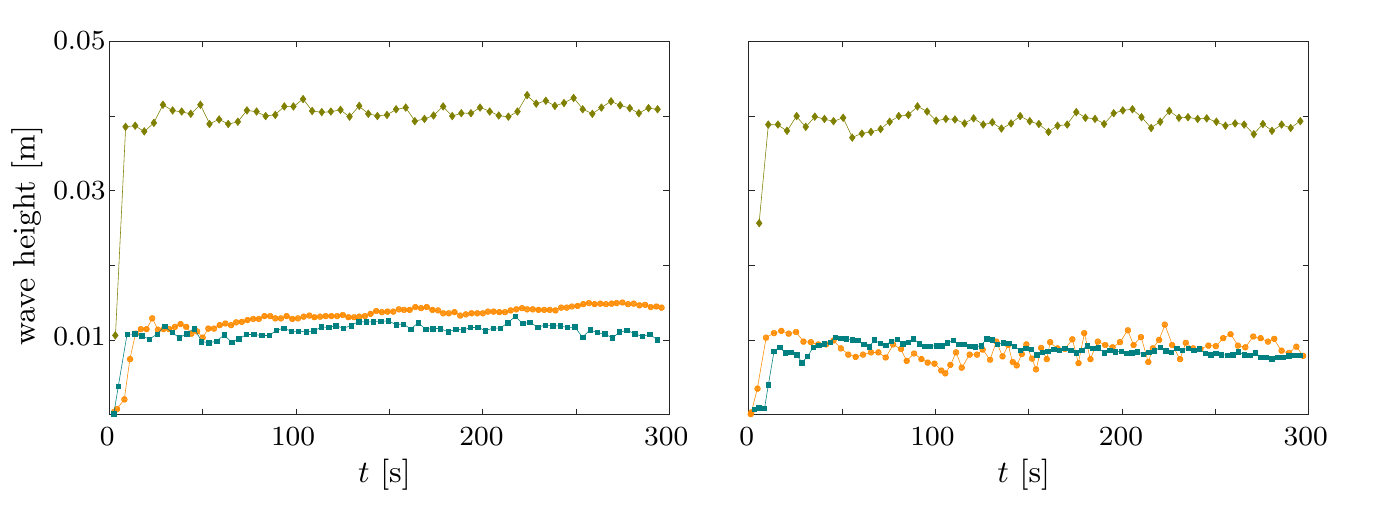}
  \end{tabular}
  \caption{Temporal variation of zero down-crossing wave height in the lee of a 5\,mm thick \textsc{pvc} floe (left-hand panel) and polypropylene floe (right).
  Incident wave has steepness $ka=0.1$ and nondimensional wavelength $L_{wave}/L_{floe}=0.56$  (teal, squares) 1 (orange, circles) and 
  {1.51} (olive, diamonds).}\label{heightP}
\end{figure}

Figs.~\ref{heightP} shows example time series of wave heights in the lee of the floe, 
for the intermediate incident wave steepness $ka=0.1$ and nondimensional wavelength $L_{wave}/L_{floe}=1{.00}$. 
The increase in wave height in the early stages of the signal is due to initial run-up as waves are generated.
No effect of the mooring transitioning from being loose to engaged is visible.
Similar behaviour is observed for lower- and higher-steepness conditions (not shown).
This does not preclude the possibility of mooring forces contributing to the generation of high-frequency transmitted wave components, for example,
by affecting the way the floe slams against the water surface.


\subsection{Transmission versus incident wave steepness and wavelength {for polypropylene floes}} 

\begin{wrapfigure}{r}{0.6\textwidth}
  \centering
  \vspace{-38pt}
  \noindent\includegraphics[width=0.35\textwidth]{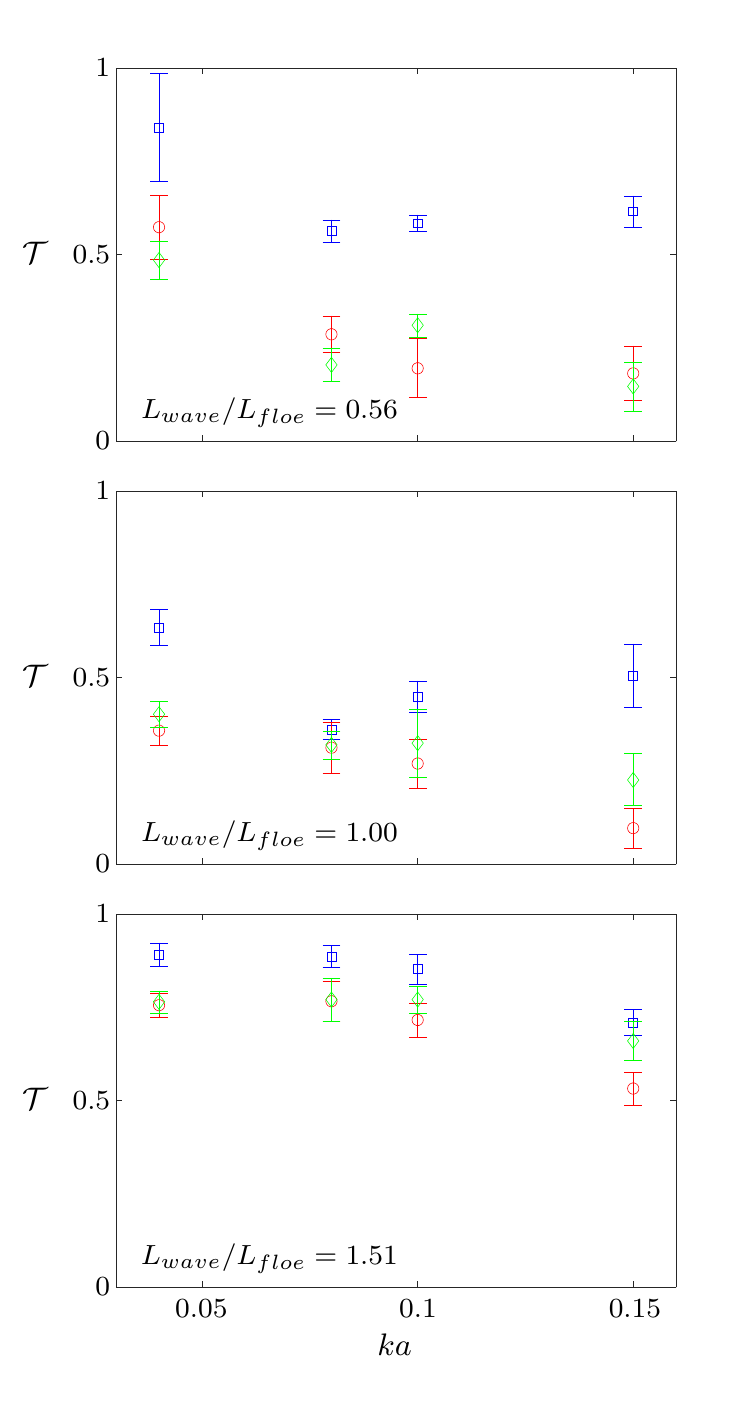}
  \caption{Transmission coefficient as a function of wave steepness {for polypropylene floes}.
  Floe thickness is $5$\,mm 
  {(blue squares)}, 10\,mm {(red circles)}, 20\,mm  {(green diamonds)}.
  {Error bars represent one standard deviation from mean.}} 
  \label{TrvsKa}
\end{wrapfigure}

The transmission coefficient for a given test is computed as 
\begin{equation}
\mathcal{T} = \frac{H}{H_{inc}}.
\end{equation}  
Here $H$ is the {mean} wave height in the lee of the floe, 
and $H_{inc}$ is the {mean} wave height of the incident wave, obtained from the control tests.  
{The transmission coefficient is presented for polypropylene floes in this section. 
Results for \textsc{pvc} floes are broadly consistent with those presented.
However, the results for polypropylene floes show the key behaviours most clearly.}  

Fig.~\ref{TrvsKa} shows the transmission coefficient, as a function of incident wave steepness.
{Error bars, denoting one standard deviation either side of the mean value, are also included.}
The results are organised according to 
{nondimensional incident wavelength (different panels) and floe thickness (different colours and symbols)}.

The transmission coefficient generally decreases with increasing incident wave steepness.
{The overall decrease from the smallest to largest steepness is greatest, approximately 0.34 to 0.39, for 
the shortest wavelength, $L_{wave}/L_{floe}=0.56$, and the two thickest floes, $h=10$\,mm and 20\,mm.
For the 5\,mm thick floes and the two shortest incident wavelengths,   $L_{wave}/L_{floe}=0.56$ and 1.00,
transmission grows with increasing steepness after initially dropping from $ka=0.04$ to 0.08.
Despite the growth, the transmission coefficient remains smaller for the largest steepness than the smallest. 
The results indicate the transmission coefficient depends on incident wave amplitude.}

\begin{wrapfigure}{r}{0.6\textwidth}
  \centering
  \noindent\includegraphics[width=0.35\textwidth]{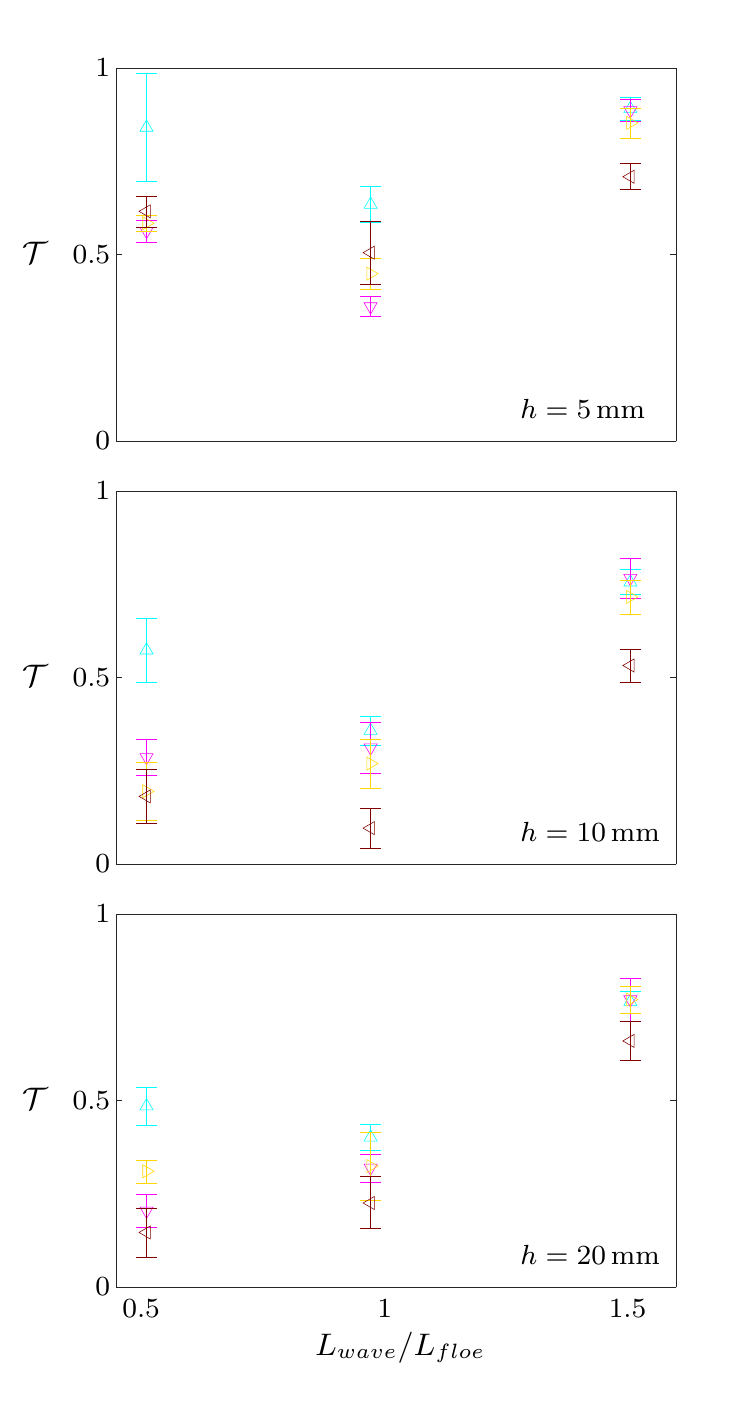}
  \caption{Transmission coefficient as a function of normalised incident wavelength {for polypropylene floes}.
  Incident wave steepness is $ka=0.04$ {(cyan, triangles up)}, 
  0.08 {(magenta down)}, 
  0.1 {(gold right)} and 0.15 {(maroon left)}.
  {Error bars represent one standard deviation from mean.}}
  \label{TrvsL}
\end{wrapfigure} 

Fig.~\ref{TrvsL} shows the transmission coefficient, as a function of nondimensional incident wavelength, $L_{wave}/L_{floe}${,
\ie a reorganised version of the transmission coefficients presented in Fig.~\ref{TrvsKa}.}
The results are organised according to
{floe thickness (different panels) and incident wave steepness (different colours and symbols).}

The transmission coefficient is, in general, minimum when the incident wavelength is equal to the floe length, and maximum for the largest incident wavelength.
As the incident wave steepness increases, 
{the transmission coefficients, generally, decrease, and the decrease is, generally, more rapid for shorter wavelengths. T}he 
minima at the intermediate incident wavelength is{, therefore,}  smoothed out.
For the thickest floe, $h=20$\,mm, 
{the minima is shallow for $ka=0.04$. 
The differential reduction with respect to wavelength as steepness increases results in a transmission coefficient that }
increases monotonically with incident wavelength for the steepest incident waves, {$ka=0.1$ and 0.15}.


\subsection{Transmission versus overwash wave height {for 10\,mm thick floes}}

\begin{figure}
  \centering 
  \noindent\includegraphics[width=0.8\textwidth]{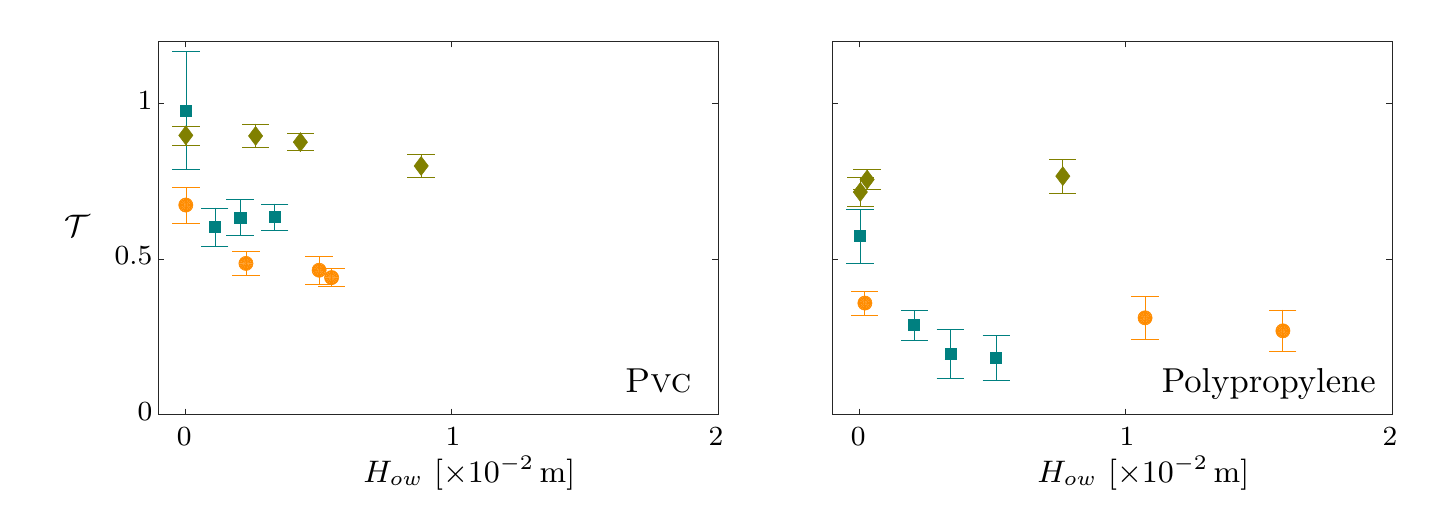}
  \caption{Transmission coefficient as a function of overwash significant wave height 
  {for 5\,mm thick \textsc{pvc} floes (left-hand panels) and polypropylene floes (right)}.
  {Nondimensional incident wavelength $L_{wave}/L_{floe}=0.56$  (teal, squares) 1.00 (orange, circles) and 
  1.51 (olive, diamonds).
   Error bars represent one standard deviation from mean.}}
  \label{TrvsOWswh}
\end{figure} 

In \S~\ref{sec:res_5mm}, it was hypothesised that wave energy dissipates in the overwash region when
the waves there become large.
The dissipation is likely to reduce the proportion of wave energy transmitted by the floe.
Fig.~\ref{TrvsOWswh} provides evidence of the effect of overwash wave height on wave transmission.
It shows the transmission coefficient as a function of the significant wave height in the overwash region, $H_{ow}$.
The significant wave height {is} calculated as four times the standard deviation of the overwash elevation, measured by the gauge on the upper surface of the floe.
The results are organised according to the plastic type {(different panels)} and nondimensional  incident wavelength
{(different colours and symbols)}.
{Results are shown for the 10\,mm thick floes, which represent the key behaviours for the other thicknesses tested.}

{The predicted trend, for the transmission coefficient to decrease as the overwash significant wave height increases, is evident.
The longest wavelength, $L_{wave}/L_{floe}=1.51$, and polypropylene floe is an exception, where the transmission coefficient is insensitive to the wave height.
A step-like decrease in transmission when the overwash wave height becomes non-zero is apparent for the shortest wavelength,  $L_{wave}/L_{floe}=0.56$.
For the \textsc{pvc} floe, the step is followed by small increments in the transmission coefficient as the wave height increases.}


\section{Summary and discussion}

An idealised experimental model of transmission of ocean waves by a solitary ice floe has been reported.
The experimental tests were conducted in a wave basin,
using regular incident waves with different wave periods, amplitudes and steepnesses. 
The incident waves ranged from gently-sloping to storm-like.
Thin plastic plates were used to model the floe. 
Two different plastics and three different thicknesses were tested. 
{A wave gauge was used to measure the wave elevation in the lee of the floe.}
Transmission coefficients were calculated from average wave heights given by the wave elevations. 
The depth of wave overwash of the floe was measured by a wave gauge, deployed in the centre of the upper surface of the floe.

{The key  findings of the data analysis are summarised and discussed below.}
\begin{enumerate}

  \item 
  The floe reduces the wave energy held in the 
  fundamental harmonic of the wave spectrum.  
  The reduction becomes more pronounced as the incident wave steepness increases.
  {This was attributed, in part, to wave energy dissipation caused by wave overwash of the floe and slamming of the floe against the water surface, both of which become stronger as the incident waves become steeper.
  Form drag around the sharp corners of the square floes, 
  is also a likely source of increased wave energy dissipation with increasing steepness.}
  
  \item
  Concomitantly, for storm-like incident waves, the floe increases the energy held in the high-frequency tail of the spectrum, 
  and the higher-order harmonics are smeared out.
  {This was attributed to the presence of high-frequency waves generated by strong overwash and slamming.}

  \item
  {Consequently, t}ransmitted waves retained regularity for gently-sloping incident waves{, but} 
  were highly irregular for storm-like incident waves. 
  
  \item
  The {proportion of wave energy transmitted, \ie the} transmission coefficient, generally, decreases as the incident wave steepness increases.
  {This is most likely a consequence of an increased proportion of wave energy being dissipated in the wave-floe interaction by the processes noted above.
  The finding implies the transmission coefficient depends on the incident wave amplitude.
  It contrasts with the predictions of existing mathematical models of wave-floe interactions, which are based on linear theory, 
  and, therefore, predict the transmission coefficient to be constant with respect to the incident wave amplitude.}
  
  \item 
  The transmission coefficient, generally, decreases, as the height of the waves in the overwash region increase.
  {The decrease was not evident in all cases, and was weak in certain cases.
  The relationship indicates wave transmission reduces due to waves breaking in the overwash region, and, hence, dissipating wave energy.}
  
\end{enumerate}

\section*{Acknowledgements}

Experiments were supported by the Small Research Grant Scheme of the School of Marine Science and Engineering of Plymouth University and performed when AA and AT were appointed at Plymouth University.
{The authors thank Peter Arber for technical support during the experiments.}
LB acknowledges funding support from the Australian Research Council (DE130101571) and the Australian Antarctic Science Grant Program (Project 4123). 
MM and AB acknowledge funding support for the U.S.\ Office of Naval Research.


\end{document}